\documentclass[10pt]{article}
\usepackage{amssymb}
\usepackage{amscd}
\usepackage{latexsym}
\usepackage{boldtensors}
\usepackage{amsmath}
\usepackage{amsthm}
\usepackage{color}
\usepackage{graphicx}
\newcommand{\Zz}{{\mathbb Z}}

\newcommand{\Rr}{{\mathbb R}}

\newcommand{\Ee}{{\mathbb E}}

\def\be{\begin{equation}}
\def\ee{\end{equation}}
\def\bea{\begin{eqnarray}}
\def\eea{\end{eqnarray}}

\def\d{{\,\rm d}}
\def\i{{\,\rm i}}

\def\0{{\bf 0}}

\def\p{{\bf p}}

\def\h2m{\frac{\hbar^2}{2m}}
\def\p0{{P_{\beta H^0_N}}}

\begin{document}
\title{
\large\bf Infinite cycles of interacting bosons}
\author{Andr\'as S\"ut\H o
\\Wigner Research Centre for Physics,  Hungarian Research Network\\
Email: suto@szfki.hu
}
\date{}
\maketitle
\thispagestyle{empty}
\begin{abstract}
\noindent
In the first-quantized description of bosonic systems permutation cycles formed by the particles play a fundamental role. In the ideal Bose gas Bose-Enstein condensation (BEC) is signaled by the appearance of infinite cycles. When the particles interact, the two phenomena may not be simultaneous, the existence of infinite cycles is necessary but not sufficient for BEC. We demonstrate that their appearance is always accompanied by a singularity
in the thermodynamic quantities
which in three and four dimensions can be as strong as a one-sided divergence of the isothermal compressibility. Arguments are presented that long-range interactions can give rise to unexpected results, such as the absence of infinite cycles in three dimensions for long-range repulsion or their presence in one and two dimensions if the pair potential has a long attractive tail.
\end{abstract}

\section{Introduction}
Feynman [F] predicted in 1953 that during Bose-Einstein condensation (BEC) the particles would form long permutation cycles. This was proven for the ideal (noninteracting) gas much later [S1, S2]. BEC of noninteracting atoms is a pure consequence of the Bose statistics. Interaction blurs the effect of the statistics and makes the proof of analogous results more difficult.
The subject of this article is not BEC. Instead,
we examine the circumstances under which infinite cycles can be present in systems of particles interacting via stable integrable pair potentials.
For reference, let us start by recalling the basic facts about the ideal Bose gas. In the absence of interaction infinite cycles appear simultaneously with  BEC,
and the phenomenon is the consequence of a kind of saturation. Let $\zeta$ denote Riemann's zeta function, $\rho$ the number density, $\beta=1/k_BT$ the inverse temperature and $\lambda_\beta=\sqrt{2\pi\hbar^2\beta/m}$ the thermal wave length associated with a particle of mass $m$. Finite cycles constitute a reservoir of particles which has infinite capacity in one and two dimensions, but in $d\geq 3$ dimensions the capacity is limited: finite cycles can be filled with particles up to $\rho=\zeta(d/2)/\lambda_\beta^d$. When the density is above this value, the surplus goes into infinite cycles that are macroscopic, i.e., each of them contains a positive fraction of the total number of particles; moreover, the infinite cycles carry the zero-momentum particles, that is, the condensate, and only them [S2]. The saturation or its absence can be read off also from the upper bound of the chemical potential $\mu$. Irrespective of the density and temperature, in the ideal Bose gas the smallest upper bound to $\mu$, i.e., its {\em supremum}, is zero, and there is saturation only if the supremum is attained, so it is a {\em maximum}. In one and two dimensions zero is only a supremum, while it is a maximum for $d\geq 3$. The difference between the two cases is manifest if first the thermodynamic limit is taken and then $\mu$ is sent to its supremum. The two limits cannot be interchanged, a fact that was not clear to Einstein's contemporaries; see e.g. [Uh] and its revision [L], [KU].

It is not obvious whether, in the case of interacting atoms, the occurrence of infinite cycles and BEC are related in the same way as in the ideal gas.
It has been shown that the two phenomena coincide in the mean-field model, and that the density of particles in infinite cycles agrees with the density of the condensate [S2], [BCMP]. This is true also for the so-called perturbed mean-field model in which the interaction is diagonal in the creation and annihilation operators, depending only on the occupation number operators $\hat N_k$ of the plane wave states [DMP]. The first deviation from the ideal gas was found in the Bose-Hubbard model with infinite-range hopping and infinite [BP], [Tth] or finite [Bo] on-site repulsion.  In this case BEC and the occurrence of infinite cycles is still simultaneous but the associated densities are not the same. The question for realistic pair potentials was addressed by Ueltschi [U1, U2] who put forward the idea that infinite cycles without BEC can occur in crystals.
There is also a closely related new field of research on models of random permutations whose Boltzmann-type probability distribution is determined by some interaction among the cycles; see e.g. [BU1], [BU2], [BUV], [EP], [AD1], [AD2], [DV]. In these models infinite cycles occur in a range of the parameters, however BEC cannot be defined independently of them.

In this paper we recall an earlier result, according to which at least the occurrence of infinite cycles is a necessary condition for BEC; as a novelty, we argue that they can appear without BEC even in one and two dimensions if the pair potential is attractive at long distances.
Since long-range attraction leads to phase separation in one dimensional classical systems [D], [J], this may be less surprising than our other finding that long-range repulsion can prevent the formation of infinite cycles and thus BEC in three dimensions. It will be seen that the appearance of infinite cycles is again the result of saturation. Finite cycles can support a density $\rho\leq \zeta_c/\lambda_\beta^d$, where $\zeta_c$ depends on $\beta$, and the surplus $\rho-\zeta_c/\lambda_\beta^d$ is carried by infinite cycles. The free energy density $f(\rho,\beta)$ is a convex function of the density [R], so $\partial f(\rho,\beta)/\partial\rho$, the chemical potential in the canonical ensemble, increases with $\rho$. It is the solution of two different equations when $\rho$ is smaller or larger than $\zeta_c/\lambda_\beta^d$; the result is a function that varies continuously, but has different analytic forms below and above $\zeta_c/\lambda_\beta^d$. In three and four dimensions the transition is of second order, $\partial^2f/\partial\rho^2$ jumps from zero to a positive value when $\rho$ passes $\zeta_c/\lambda_\beta^d$ increasingly, implying a similar jump of the derivative of the pressure with respect to $\rho$ and giving rise to a one-sided divergence of the isothermal compressibility. If the limit of $\zeta_c$ is finite when $\beta$ goes to infinity, at zero temperature all the particles are in infinite cycles, even though BEC cannot be complete.

In the following we discuss the formulas and what we expect when only finite cycles are present, draw partly conjectural conclusions about infinite cycles, and make some final comments.

\section{Finite cycles in the infinite system}

The analysis is based on the path integral representation of the partition function. A detailed treatment of the path integral formalism in Quantum Statistical Mechanics can be found in Ginibre's Lecture Notes [G]. Other useful sources are the books [Si], [Ro] and [LHB]. Consider $N$ identical bosons on the $d$-dimensional torus $\Lambda=(-L/2,L/2]^d$ (hypercube of side $L$ with periodic boundary conditions). Let $X=(x_1,\dots,x_N)\in\Lambda^{N}$, then the canonical partition function is
\[
Q_{N,L}=\frac{1}{N!}\sum_{\pi\in S_N}\int_{\Lambda^{N}}\d X\int W^\beta_{X,\pi X}(\d\Omega)e^{-\beta U(\Omega)}.
\]
Here $S_N$ is the group of permutations,
$
\pi X=(x_{\pi(1)},\dots,x_{\pi(N)}),
$
and $W^\beta_{X,\pi X}(\d\Omega)$ is the Brownian bridge measure on the torus for $N$-particle trajectories $\Omega(t)=(\omega_1(t),\dots,\omega_N(t))$, $t\in [0,\beta]$, such that $\Omega(0)=X$ and $\Omega(\beta)=\pi X$.
Furthermore,
\[
\beta U(\Omega)= \sum_{1\leq j<k\leq N}\int_0^\beta u_L(\omega_k(t)-\omega_j(t)) \d t
\]
where $u_L$ is the periodized pair potential,
\[
u_L(x)=\sum_{z\in\Zz^d}u(x+Lz),\quad u(x)=u(-x).
\]
To be clear,  $\omega_i: [0,\beta]\to \Rr^d$,
\be\label{defW}
W^\beta_{X,\pi X}(\d\Omega)=\prod_{i=1}^N W^\beta_{x_i,x_{\pi(i)}}(\d\omega_i),
\qquad
W^\beta_{xy}(\d\omega)=\sum_{z\in\Zz^d}P^\beta_{x,y+Lz}(\d\omega),
\ee
and $P^{\beta}_{x,y+Lz}(\d\omega)$ is the Brownian bridge measure for trajectories $\omega$ in $\Rr^d$ that start in $x$ at $t=0$ and end in $y+Lz$ at $t=\beta$. In
other words, the one-particle trajectories are not on the torus but in $\Rr^d$, and it is the integrand $\Phi(\Omega)$ that satisfies $\Phi(\Omega)=\Phi(\Omega+(Lz_1,\dots,Lz_N))$ for any $z_1,\dots,z_N\in\Zz^d$.

Each $\Omega$ breaks up into permutation cycles, and by distinguishing the cycle $\omega$ that contains particle no.1, $Q_{N,L}$ can be rewritten in a form close to a recurrence relation,
\begin{eqnarray}\label{QNL}
Q_{N,L}
=\frac{1}{N}\sum_{n=1}^N  \int_\Lambda\!\! \d x\!\! \int W^{n\beta}_{xx}(\d\omega)e^{-\beta U(\omega)} Q_{N-n,L}(\omega)
=\frac{1}{\rho}\sum_{n=1}^N  \int W^{n\beta}_{00}(\d\omega)e^{-\beta U(\omega)} Q_{N-n,L}(\omega)
\hspace{1cm}
\end{eqnarray}
where $\rho=N/L^d$.
The second form follows from translation invariance,
\[
\beta U(\omega)=\sum_{0\leq j< k\leq n-1}\int_0^\beta u_L(\omega(k\beta+t)-\omega(j\beta+t))\d t,
\]
and
\begin{eqnarray*}
Q_{N-n,L}(\omega)=\frac{1}{(N-n)!}\sum_{\pi\in S_{N-n}}
\int_{\Lambda^{N-n}}\d X
\int W^\beta_{X,\pi X}(\d\Omega)e^{-\beta U(\Omega)} e^{-\beta U(\omega,\Omega)}.
\end{eqnarray*}
In the above formula $\Omega$ is a $N-n$-particle trajectory and
\be\label{UomegaOmega}
\beta U(\omega,\Omega)
=\sum_{j=0}^{n-1} \sum_{k=1}^{N-n}\int_0^\beta u_L(\omega_k(t)-\omega(j\beta+t)) \d t.
\ee
Without $e^{-\beta U(\omega,\Omega)}$, $Q_{N-n,L}(\omega)$ is just $Q_{N-n,L}$.
We will analyze the equation
\[
\rho=\!\sum_{n=1}^N\!  \int W^{n\beta}_{00}(\d\omega) e^{-\beta U(\omega)} \frac{Q_{N-n,L}(\omega)}{Q_{N,L}}
= \sum_{n=1}^N \rho^{N,L}_n
\]
in the thermodynamic limit. $\rho^{N,L}_n$ is deduced from the previous equality where $\rho^{N,L}_n/\rho$ corresponds to the probability according to the canonical Gibbs distribution that particle no.1 is in a cycle of length $n$. Because the particles are indistinguishable, $\rho^{N,L}_n$ is also the partial density carried by $n$-cycles.

We cut the sum into two, from 1 to $M$ and from $M+1$ to $N$ and take $\lim_{M\to\infty}\lim_{N,L\to\infty, N/L^d=\rho}$. In the sum up to $M$ the thermodynamic limit can be executed under the summation sign, therefore
\begin{eqnarray*}
\rho=\sum_{n=1}^\infty \lim_{N,L\to\infty, N/L^d=\rho}\rho^{N,L}_n
+\lim_{M\to\infty}\lim_{N,L\to\infty, N/L^d=\rho}\sum_{n=M+1}^N\rho^{N,L}_n.
\end{eqnarray*}
The infinite sum is the contribution of the finite cycles in the infinite system, the second term is that of the infinite cycles. Let us suppose that the latter is zero, then the equation to investigate is
\be\label{only-finite-cycles}
\rho=\sum_{n=1}^\infty \lim_{N,L\to\infty, N/L^d=\rho}\rho^{N,L}_n=\sum_{n=1}^\infty \rho_n,
\ee
and we would like to see how can it fail if $\rho\lambda_\beta^d$ is large enough. However, first we recall from [Ue], [S3] that Eq.~(\ref{only-finite-cycles}) implies the absence of BEC.

O. Penrose and Onsager [PO] postulated BEC of interacting bosons as the increase of the largest eigenvalue of the one-particle reduced density matrix $\sigma^{N,L}_1$ proportionally to $N$.
This extends the original definition that BEC is the presence of zero-momentum particles in numbers proportional to $N$.
Indeed, on a hypercube of side $L$ with periodic boundary conditions
\[
\sigma^{N,L}_1=\sum_{k\in (2\pi/L)\Zz^d} N_{k} |k\rangle\langle k|.
\]
Here $| k\rangle\langle k|$ is the orthogonal projection to the one-dimensional subspace spanned by $L^{-d/2}e^{\i k\cdot x}$, and
$ N_k$ is the expected number of particles in that state. The Banach-space analogue of the Perron-Frobenius theorem [KR] implies that $ N_{0}$ is the largest eigenvalue with eigenvector identically equal to $L^{-d/2}$.
 Let $\langle x|\sigma^{N,L}_1|y\rangle$ denote the integral kernel of $\sigma^{N,L}_1$ (the off-diagonal correlation function), then
\[
\rho^{N,L}_0=\frac{N_{0}}{L^d}=\frac{1}{{L^d}}\int_\Lambda \langle x|\sigma^{N,L}_1|0\rangle \d x.
\]
Utilizing this identity, one can prove the inequality
\begin{eqnarray*}
\rho^{N,L}_0\leq \sum_{n=1}^N \rho^{N,L}_n \frac{1}{L^d}\int_\Lambda\exp\left\{-\frac{\pi x^2}{n\lambda_\beta^2}\right\}\d x.
\end{eqnarray*}
In the absence of infinite cycles the thermodynamic limit can be taken term by term, resulting zero for each term.

Now we start the  analysis of Eq.~(\ref{only-finite-cycles}).
We introduce three probability densities,
\begin{eqnarray*}
m_{n,L}(\omega)\equiv \frac{1}{\int W^{n\beta}_{00}(\d\omega')},\quad
m^u_{n,L}(\omega)=\frac{e^{-\beta U(\omega)}}{\int W^{n\beta}_{00}(\d\omega')e^{-\beta U(\omega')}},\quad
M_{N-n,L}(\Omega)=\frac{e^{-\beta U(\Omega)}}{(N-n)!Q_{N-n,L}}.
\end{eqnarray*}
The first two is defined on single-particle trajectories starting and ending in 0 in the time interval $[0,n\beta]$, the first being the uniform distribution on this set. The third probability measure is on the set of $N-n$-particle trajectories that start in some $X\in\Lambda^{N-n}$ at time 0 and end in $\pi X$ at time $\beta$, where $\pi$ is any permutation of $1,\dots,N-n$.
With them $\rho^{N,L}_n$ can be divided into four factors,
\be\label{rhoNLfactors}
\rho^{N,L}_n
=
\left(\int W^{n\beta}_{00}(\d\omega)\right)\frac{Q_{N-n,L}}{Q_{N,L}}
\,\Ee\left[e^{-\beta U(\cdot)}\right]_{m_{n,L}}
\ \Ee\left[e^{-\beta U(\cdot,\cdot)}\right]_{m^u_{n,L}\times M_{N-n,L}},
\ee
where
\begin{eqnarray*}
\Ee\left[e^{-\beta U(\cdot)}\right]_{m_{n,L}}
=\int W^{n\beta}_{00}(\d\omega) e^{-\beta U(\omega)}m_{n,L}(\omega)
\end{eqnarray*}
and
\begin{eqnarray*}
\Ee\left[e^{-\beta U(\cdot,\cdot)}\right]_{m^u_{n,L}\times M_{N-n,L}}
=\int W^{n\beta}_{00}(\d\omega)m^u_{n,L}(\omega)
 \sum_{\pi\in S_{N-n}}\int_{\Lambda^{N-n}}\d X\int W^\beta_{X,\pi X}(\d\Omega)M_{N-n,L}(\Omega)e^{-\beta U(\omega,\Omega)}.
\end{eqnarray*}
The thermodynamic limit of the first two factors of $\rho^{N,L}_n$ is easily obtained. The first is independent of the density, and with
\[
\int P^{t}_{xy}(\d\omega)=\lambda_t^{-d}e^{-\pi(x-y)^2/\lambda_t^2},\qquad \lambda_t=\sqrt{2\pi\hbar^2t/m}
\]
its limit is
\be\label{WtoP}
\lim_{L\to\infty}\int W^{n\beta}_{00}(\d\omega)
=\frac{1}{\lambda_\beta^d n^{d/2}}\left(1+
\lim_{L\to\infty}\sum_{0\neq z\in\Zz^d}e^{-L^2 z^2/n\lambda_\beta^2}\right)
=\frac{1}{\lambda_\beta^d n^{d/2}},
\ee
cf. Eq.~(\ref{defW}).

Concerning the second factor in (\ref{rhoNLfactors}), let us recall that the free energy $F_{N,L}$ of $N$ particles in the hypercube of side $L$ and the free energy per unit volume $f_{N,L}=F_{N,L}/L^d$ are related to $Q_{N,L}$ through the formula
\[
Q_{N,L}=e^{-\beta F_{N,L}}=e^{-\beta L^df_{N,L}},
\]
so
\[
\frac{{Q}_{N-1,L}}{{Q}_{N,L}}=e^{\beta (F_{N,L}-F_{N-1,L})}
=\exp\left\{\beta\frac{f_{N,L}-f_{N-1,L}}{N/L^d-(N-1)/L^d}\right\}\to \exp\left\{\beta\frac{\partial {f}(\rho,\beta)}{\partial \rho}\right\}\quad(N,L\to\infty,N/L^d=\rho).
\]
Therefore
\be\label{chempot}
\lim_{N,L\to\infty, N/L^d=\rho} \frac{{Q}_{N-n,L}}{{Q}_{N,L}}
=\!\!\lim_{N,L\to\infty, N/L^d=\rho}\prod_{i=1}^n \frac{{Q}_{N-i,L}}{{Q}_{N-i+1,L}}
= \left(e^{\beta \partial {f}(\rho,\beta)/\partial \rho}\right)^n,
\ee
where
\[
f(\rho,\beta)=\lim_{N,L\to\infty, N/L^d=\rho}f_{N,L}
\]
is the free energy density and
$\partial f(\rho,\beta)/\partial\rho$ is the chemical potential. The formulas are valid for stable pair potentials, and because $f(\rho,\beta)$ is a convex function of $\rho$, $\partial f(\rho,\beta)/\partial\rho$ exists for almost all $\rho$.

The third factor is also independent of the density,
\be\label{Eeto-betaU}
\lim_{L\to\infty}\Ee\left[e^{-\beta U(\cdot)}\right]_{m_{n,L}}
=\frac{\int P^{n\beta}_{00}(\d\omega) e^{-\beta U(\omega)}}{\int P^{n\beta}_{00}(\d\omega')}
=\Ee\left[e^{-\beta U(\cdot)}\right]_{P^{n\beta}_{00}}
\ee
where $U(\omega)$ is defined with $u$ instead of $u_L$. For an integrable $u$ in any dimension one has [S3]
\be\label{E[U]}
\Ee[U(\cdot)]_{P^{n\beta}_{00}}=\frac{\int P^{n\beta}_{00}(\d\omega) U(\omega)}{\int P^{n\beta}_{00}(\d\omega')}
=\frac{n}{2}\sum_{k=1}^{n-1}\alpha_{n,k}^{d/2}\int u(x)e^{-\pi\alpha_{n,k}x^2}\d x
\ee
where
$
\alpha_{n,k}=\lambda_\beta^{-2}\left[k^{-1}+(n-k)^{-1}\right].
$
Bounding the right-hand side above and utilizing Jensen's inequality, for $d\geq 3$
\bea\label{lower-to-exp-beta U(omega)}
\Ee\left[e^{-\beta U(\cdot)}\right]_{P^{n\beta}_{00}}
\geq
e^{-\beta\Ee[U(\cdot)]_{P^{n\beta}_{00}}}
\geq
e^{-2^{d/2-1}\zeta(d/2)\frac{\beta\|u\|_1}{\lambda_\beta^d}n\left(1-\frac{1}{\zeta(d/2)}\sum_{k=n/2+1}^{\infty}\frac{1}{k^{d/2}}\right)}.
\eea
Here $\|u\|_1=\int |u(x)|\d x$. The expected value (\ref{Eeto-betaU}) can be written as
\be\label{epsilon_n}
 \Ee\left[e^{-\beta U(\cdot)}\right]_{P^{n\beta}_{00}}
=e^{-n\beta\epsilon_n(\beta)},
\ee
implicitly defining $\epsilon_n(\beta)$.
By stability [R], $\epsilon_n(\beta)\geq -B$ with some $B\geq 0$, and from (\ref{lower-to-exp-beta U(omega)}) one infers
\begin{eqnarray}\label{upper_to_epsilon_n}
\epsilon_n(\beta)
\leq n^{-1}\Ee[U(\cdot)]_{P^{n\beta}_{00}}
\leq 2^{d/2-1}\zeta(d/2)\frac{\|u\|_1}{\lambda_\beta^d}\left[1-\frac{\sum_{k=n/2+1}^{\infty}\frac{1}{k^{d/2}}}{\zeta(d/2)}\right].
\end{eqnarray}

Assuming that $\rho_n=\lim_{N,L\to\infty, N/L^d=\rho}\rho^{N,L}_n$ exists, the fourth factor of $\rho^{N,L}_n$ also has a limit. Let
\begin{eqnarray*}
\lim_{N,L\to\infty, N/L^d=\rho}\Ee\left[e^{-\beta U(\cdot,\cdot)}\right]_{m^u_{n,L}\times M_{N-n,L}}
=\Ee\left[e^{-\beta U(\cdot,\cdot)}\right]_{\nu^u_n\times G(\rho,\beta)}
\end{eqnarray*}
and, in analogy with $\epsilon_n(\beta)$, define $E_n(\rho,\beta)$ via the equation
\be\label{E_n}
\Ee\left[e^{-\beta U(\cdot,\cdot)}\right]_{\nu^u_n\times G(\rho,\beta)}
=e^{-n\beta E_n(\rho,\beta)}.
\ee
Above, $\nu^u_n$ is the probability measure of density
\[
\lim_{L\to\infty}m^u_{n,L}(\omega)
=\frac{e^{-\beta U(\omega)}}{\int P^{n\beta}_{00}(\d\omega')e^{-\beta U(\omega')}}
\]
and $G(\rho,\beta)$ is the infinite volume Gibbs measure generated by periodic boundary conditions.

Due to translation invariance on the torus the expected value of $U(\omega,\Omega)$ is easily obtained. Its taking involves integration with respect to the initial points $\omega_k(0)$ ($k~=0,\dots,N-n$) in $\Lambda$. Any additional shift or average over all possible shifts of $\omega_k$ is therefore redundant but useful. By first integrating over a global shift $y$ we obtain
\[
\int_\Lambda u_L(\omega_k(t)-\omega(j\beta+t)+y)\d y
=\sum_{z\in \Zz^d}\int_\Lambda u(\omega_k(t)-\omega(j\beta+t)+y+Lz)\d y
=\int_{\Rr^d} u(x)\d x =I_u.
\]
The result is independent of $\omega$, $\omega_k$ and $t$, so subsequent averages over their sets do not change anything. The sums in Eq.~(\ref{UomegaOmega}) with respect to $j$ and $k$ provide a factor $n(N-n)$ which, with the remaining division with $L^d$ to complete the average over the global shift gives
\begin{eqnarray*}
\Ee\left[U(\cdot,\cdot)\right]_{m^u_{n,L}\times M_{N-n,L}}
=\frac{n(N-n)}{L^d} I_u
\end{eqnarray*}
and
\[
\Ee\left[U(\cdot,\cdot)\right]_{\nu^u_n\times G(\rho,\beta)}
=n\rho I_u.
\]
By Jensen's inequality,
\begin{equation}\label{lower-to-exp-U(om,Om)}
e^{-n\beta E_n(\rho,\beta)}=\Ee\left[e^{-\beta U(\cdot,\cdot)}\right]_{\nu^u_n\times G(\rho,\beta)}
\geq e^{-\beta \Ee\left[U(\cdot,\cdot)\right]_{\nu^u_n\times G(\rho,\beta)}}
= e^{-n\beta \rho I_u},
\end{equation}
so
\be\label{upper-to-E_n(rho,beta)}
E_n(\rho,\beta)\leq \rho I_u.
\ee
Because of stability $E_n(\rho,\beta)$ is bounded below by a constant; if $u\geq 0$ then $E_n(\rho,\beta)\geq 0$.

The thermodynamic limit $\rho_n$ of $\rho^{N,L}_n$ is obtained by inserting the respective limits (\ref{WtoP}), (\ref{chempot}), (\ref{epsilon_n}) and (\ref{E_n}) of the four factors into Eq.~(\ref{rhoNLfactors}). Introducing ${\cal E}_n(\rho,\beta)=\epsilon_n(\beta)+E_n(\rho,\beta)$,
\[
\rho_n=\frac{e^{n\beta[\partial f/\partial\rho-{\cal E}_n(\rho,\beta)]}}{n^{d/2}\lambda_\beta^d}.
\]
With the definition
\be\label{A_rho_beta}
A_{\rho,\beta}(\mu)=\sum_{n=1}^\infty \frac{e^{n\beta[\mu-{\cal E}_n(\rho,\beta)]}}{n^{d/2}}
\ee
equation (\ref{only-finite-cycles}) becomes
\be\label{A_rho_beta=}
A_{\rho,\beta}(\mu)= \rho\lambda_\beta^d,
\ee
to be solved for $\mu=\partial {f}(\rho,\beta)/\partial \rho$.

Being in bounded intervals, both $\epsilon_n(\beta)$ and $E_n(\rho,\beta)$
and thus ${\cal E}_n(\rho,\beta)$
have a finite limit when $n$ goes to infinity (we disregard the possibility of more than one limit point). Let $\epsilon(\beta)$, $E(\rho,\beta)$ and ${\cal E}(\rho,\beta)=\epsilon(\beta)+E(\rho,\beta)$ denote the respective limits. We shall see that what counts for the solubility of Eq.~(\ref{A_rho_beta=}) is not ${\cal E}(\rho,\beta)$ but the way ${\cal E}_n(\rho,\beta)$ tends to it. If $u\geq 0$ then ${\cal E}_n(\rho,\beta)>0$, and we expect ${\cal E}_{n+1}\geq {\cal E}_n$ because in
\[
\beta
U(\omega)=\sum_{0\leq j<k\leq n}\int_0^\beta u(\omega(k\beta+t)-\omega(j\beta+t))\d t
=\frac{1}{2}\sum_{j=0}^{n}\sum_{k\in\{0,\dots.n\}, k\neq j}\int_0^\beta u(\omega(k\beta+t)-\omega(j\beta+t))\d t,
\]
written for a cycle of $n+1$ particles, both sums have one more term: compared to a $n$-cycle, all the particles interact repulsively with one more particle.
Although ${\cal E}_{n+1}\geq {\cal E}_n$ may be true for sufficiently large $n$, for all $n$ we can prove only less.

\vspace{5pt}
\noindent
{\em Lemma.} If $u\geq 0$ then
for any $n\geq 1$
\be\label{Mn+1-Mn}
(n+1){\cal E}_{n+1}(\rho,\beta)\geq n{\cal E}_n(\rho,\beta).
\ee

\vspace{5pt}
\noindent
{\em Proof.} Let
\begin{eqnarray*}
\Ee\left[e^{-\beta U(\omega,\cdot)}\right]_{M_{N-n,L}}=
\frac{1}{(N-n)!Q_{N-n,L}}\sum_{\pi\in S_{N-n}}
\int_{\Lambda^{N-n}}\d X\int W^\beta_{X,\pi X}(\d\Omega)e^{-\beta U(\Omega)} e^{-\beta U(\omega,\Omega)},
\end{eqnarray*}
\bea\label{e-U(om,Om)-after-limit}
\Ee\left[e^{-\beta U(\omega,\cdot)}\right]_{G(\rho,\beta)}
=\lim_{N,L\to\infty, N/L^d=\rho}\Ee\left[e^{-\beta U(\omega,\cdot)}\right]_{M_{N-n,L}}
\eea
and
\[
b_n(x)=\int P^{n\beta}_{0x}(\d\omega)e^{-\beta U(\omega)}\Ee\left[e^{-\beta U(\omega,\cdot)}\right]_{G(\rho,\beta)}.
\]
Then,
by comparison with Eqs.~(\ref{epsilon_n}) and (\ref{E_n}),
\be\label{b_n-M_n}
\frac{b_n(0)}{\int P^{n\beta}_{00}(\d\omega)}=e^{-n\beta {\cal E}_n(\rho,\beta)}.
\ee
Now
\begin{eqnarray*}
b_{n+1}(0)&=&\int\d x\int P^{n\beta}_{0x}(\d\omega_1)e^{-\sum_{0\leq j<k\leq n-1}\int_0^\beta u(\omega_1(k\beta+t)-\omega_1(j\beta+t))\d t}
\\
&&\int P^{\beta}_{x0}(\d\omega_2)e^{-\sum_{j=0}^{n-1}\int_0^\beta u(\omega_2(t)-\omega_1(j\beta+t))\d t}
\Ee\left[e^{-\beta U(\omega_1\circ\,\omega_2,\cdot)}\right]_{G(\rho,\beta)}
\\
&\leq&\int\d x\int P^{\beta}_{x0}(\d\omega_2)\int P^{n\beta}_{0x}(\d\omega_1)e^{-\beta U(\omega_1)}\Ee\left[e^{-\beta U(\omega_1,\cdot)}\right]_{G(\rho,\beta)}
=\int\d x\frac{e^{-\pi x^2/\lambda_\beta^2}}{\lambda_\beta^d} b_n(x).
\end{eqnarray*}
Above, $\omega_1\circ\,\omega_2$ is the concatenation of $\omega_1$ and $\omega_2$. By symmetry, $\nabla b_n(x)|_{x=0}=0$.
In [S3] we proved that $b_n(x)$ has its global maximum at $x=0$,
\[
b_n(x)\leq e^{-\pi x^2/\lambda_{n\beta}^2} b_n(0).
\]
Recalling that $\lambda_{n\beta}=\sqrt{n}\lambda_\beta$ and $\int_{\Rr^d} e^{-ax^2}\d x=(\pi/a)^{d/2}$,
evaluating the integral with the upper bound gives $b_{n+1}(0)\leq (\frac{n}{n+1})^{d/2} b_n(0)$ which with (\ref{b_n-M_n}) gives
\[
e^{-(n+1)\beta {\cal E}_{n+1}(\rho,\beta)}\leq e^{-n\beta {\cal E}_{n}(\rho,\beta)}
\]
and therefore (\ref{Mn+1-Mn}). $\blacksquare$

\vspace{5pt}
With (\ref{Mn+1-Mn}) one obtains
\be\label{diff-cal-E}
{\cal E}_{n+1}- {\cal E}_n=\frac{1}{n}\left(-{\cal E}_{n+1}+(n+1){\cal E}_{n+1}-n{\cal E}_n\right)\geq -\frac{1}{n}{\cal E}_{n+1}.
\ee
In the proof of the Lemma the energy of $\omega_2$ in the field of $\omega_1$ and $\Omega$ was bounded below by zero. For typical trajectories this energy is of order 1, therefore it adds a positive term of order $1/n$ to the right-hand-side of the inequality (\ref{diff-cal-E}). Because ${\cal E}_{n+1}- {\cal E}_n\to 0$, the lemma does not exclude ${\cal E}_{n+1}-{\cal E}_n< 0$ for $u\geq 0$ even with this correction. Note that ${\cal E}_{n+1}- {\cal E}_n=o(1/n)$ if it is sign-keeping, otherwise $\lim_{n\to\infty}{\cal E}_n$ could not be finite.

\section{Infinite cycles}

With
\[
D_n(\rho,\beta)={\cal E}(\rho,\beta)-{\cal E}_n(\rho,\beta)
\]
equation (\ref{A_rho_beta=}) takes the form
\be\label{A_rho_beta=bis}
A_{\rho,\beta}(\mu)\equiv\sum_{n=1}^\infty \frac{e^{n\beta D_n(\rho,\beta)]}}{n^{d/2}}e^{n\beta[\mu-{\cal E}(\rho,\beta)]}=\rho\lambda_\beta^d.
\ee
Let
\[
\delta_n(\beta)=\epsilon(\beta)-\epsilon_n(\beta),\quad \Delta_n(\rho,\beta)=E(\rho,\beta)-E_n(\rho,\beta),
\]
then
$$D_n(\rho,\beta)=\delta_n(\beta)+\Delta_n(\rho,\beta).$$
By definition, both $\delta_n(\beta)$ and $\Delta_n(\rho,\beta)$ go to zero as $n$ goes to infinity. Moreover,
\[
D_n-D_{n+1}={\cal E}_{n+1}- {\cal E}_n.
\]
It is the sequence $D_n$ that decides about the existence of infinite cycles.

\subsection{Infinite cycles avoided}

A prerequisite of the solubility of Eq.~(\ref{A_rho_beta=bis}) is that the infinite sum is convergent. Because $nD_n(\rho,\beta)=o(n)$, this is true for $\mu< {\cal E}(\rho,\beta)$ and fails for $\mu>{\cal E}(\rho,\beta)$. So the supremum of $\mu$ is ${\cal E}(\rho,\beta)$. It is a maximum if
\[
\zeta_c(\rho,\beta)=A_{\rho,\beta}({\cal E}(\rho,\beta))=\sum_{n=1}^\infty \frac{e^{n\beta D_n(\rho,\beta)}}{n^{d/2}}
\]
is finite, and this is necessary for the occurrence of infinite cycles. To see why, assume, for example, that at a given density for all temperatures $nD_n(\rho,\beta)\to\infty$
as $n\to\infty$, and its divergence is not too slow, making $\zeta_c(\rho,\beta)=\infty$. Then, increasing $\beta$ and varying $\mu$ so that $\mu-{\cal E}(\rho,\beta)<0$ raises to zero, $A_{\rho,\beta}(\mu)$ is convergent and goes continuously to infinity. Therefore, $A_{\rho,\beta}(\mu)=\rho\lambda_\beta^d$ can be solved for $\mu$, regardless of the size of $\rho\lambda_\beta^d$. This is analogous to how the BEC of the noninteracting gas is avoided for $d=1,2$ due to $\sum_{n=1}^\infty n^{-d/2}=\infty$.
Below we present a possible realization of this scenario in three dimensions for long-range repulsive interactions.

More generally, let $u$ be a long-range potential such that for some $R>0$, $d<\eta_2\leq\eta_1<\infty$, and $0<b_1\leq b_2$,
\be\label{upper-lower-on-u}
b_1|x|^{-\eta_1}\leq |u(x)|\leq b_2|x|^{-\eta_2}\quad  \mbox{if $|x|>R$}.
\ee
$\eta_2>d$ is necessary for the integrability of $u$ and the convergence of the infinite sum $u_L$.
For interactions satisfying (\ref{upper-lower-on-u}) we make the following hypothesis: If $k$ is sufficiently large and $\omega$ is a $P^{(k+1)\beta}_{00}$-typical trajectory, then
\be\label{Delta-epsilon-estimate}
\epsilon_{k+1}(\beta)-\epsilon_k(\beta)\sim u(\omega((k+1)\beta/2)).
\ee
The intuition behind this formula is that $|\omega((k+1)\beta/2)|\sim \sqrt{(k+1)/2}\lambda_\beta$ is the largest expected distance between two points of $\omega$, so when the number of particles is increased by one, each particle in the cycle experiences a change of energy $\sim u(\omega((k+1)\beta/2))$. Here and below $\sim$ means an order-of-magnitude agreement.

Suppose that
\be\label{ek+1-ek}
\sum_{k=1}^\infty |\epsilon_k(\beta)-\epsilon_{k+1}(\beta)|<\infty,
\ee
then
\be\label{-delta}
\delta_n(\beta)=-\sum_{k=n}^\infty \left[\epsilon_k(\beta)-\epsilon_{k+1}(\beta)\right].
\ee
Indeed, for any $M>n$
\begin{eqnarray*}
\sum_{k=n}^\infty (\epsilon_k-\epsilon_{k+1})
&=&\epsilon_n-\epsilon_M+\sum_{k=M}^\infty (\epsilon_k-\epsilon_{k+1})
\\
&\rightarrow& \epsilon_n-\epsilon=-\delta_n
\end{eqnarray*}
as $M$ goes to infinity.

Let now $u\geq 0$ satisfy (\ref{upper-lower-on-u}). Then, for $k\geq n$ sufficiently large (\ref{Delta-epsilon-estimate}) implies
\[
b_1(\sqrt{k/2}\lambda_\beta)^{-\eta_1}\lesssim \epsilon_{k+1}(\beta)-\epsilon_k(\beta)\lesssim b_2(\sqrt{k/2}\lambda_\beta)^{-\eta_2}.
\]
If $\eta_2>2$, equations (\ref{ek+1-ek}) and (\ref{-delta}) are true, so
\[
n^{1-\eta_1/2}\lesssim \delta_n(\beta)\lesssim n^{1-\eta_2/2}.
\]
In three dimensions $\eta_2>3$; however, if $\eta_1<4$, $n\delta_n(\beta)\gtrsim n^{2-\eta_1/2}\to\infty$ when $n\to\infty$. Assuming that the sequence $n\Delta_n(\rho,\beta)$ does not go to minus infinity at the same rate or faster, $\zeta_c(\rho,\beta)=\infty$, and we obtain the following.

\vspace{5pt}\noindent
{\em Conjecture 1.} Let $d=3$. Suppose that $u\geq 0$ satisfies the bounds (\ref{upper-lower-on-u}) with $3<\eta_2\leq\eta_1<4$. Then, for all $\rho, \beta<\infty$ there are no infinite cycles and, therefore, no BEC.

\vspace{5pt}
There is a theoretically possible, more sophisticated reason why infinite cycles can be missing: $\zeta_c(\rho,\beta)$, even though finite, increases with $\rho$ or $\beta$ in such a way that $\zeta_c(\rho,\beta)\geq\rho\lambda_\beta^d$, making Eq.~(\ref{A_rho_beta=bis}) soluble for arbitrarily large values of $\rho\lambda_\beta^d$. This could be the result of either $\beta\delta_n(\beta)$ increasing with $\beta$ or $\beta\Delta_n(\rho,\beta)$ increasing with $\rho$ or $\beta$. The first can be ruled out, at least for repulsive interactions:
If $u\geq 0$, for $d\geq 3$ we have $0\leq\epsilon_n(\beta)\leq c\|u\|_1/\lambda_\beta^d$, cf. (\ref{upper_to_epsilon_n}), and thus
\[
\beta \epsilon_n(\beta),\ \beta \epsilon(\beta), \ \beta\delta_n(\beta)\to 0 \quad \mbox{if $\beta\to\infty$}.
\]
The upper bound (\ref{upper-to-E_n(rho,beta)}) does not allow us to draw a firm conclusion on $\beta\Delta_n(\rho,\beta)$. Nevertheless, its independence with respect to $n$ suggests that $\Delta_n(\rho,\beta)\equiv 0$, and intuition supports this.
$\Ee\left[e^{-\beta U(\cdot,\cdot)}\right]_{m^u_{n,L}\times M_{N-n,L}}$ is not a purely exponential function of $n$ because in $U(\omega,\Omega)$ the sum over $k$ runs from 1 to $N-n$. After the thermodynamic limit the average (\ref{e-U(om,Om)-after-limit}) can be independent of $\omega$ and purely exponential in $n$, because the single-particle distribution in the Gibbs state $G(\rho,\beta)$ is uniform, and $U(\omega,\Omega)$ acts as an external field exerted by the particles of $\Omega$ on those of $\omega$. Because $\delta_1(\beta)=\epsilon(\beta)$, $\Delta_n(\rho,\beta)\equiv 0$ implies $\zeta_c(\beta)\geq e^{\beta\epsilon(\beta)}$ and thus the absence of infinite cycles and BEC if $\rho\lambda_\beta^d\leq e^{\beta\epsilon(\beta)}$. Now $\epsilon(\beta)\geq 0$ for $u\geq 0$, therefore, under the hypothesis $\Delta_n(\rho,\beta)\equiv 0$, infinite cycles and BEC are missing for $\rho\lambda_\beta^d\leq 1$.

\subsection{Meeting  infinite cycles}

Even if  $\Delta_n(\rho,\beta)\neq 0$, there is an asymmetry in the $\rho$- and $\beta$-dependence of $\zeta_c(\rho,\beta)$. The situation is clearer if $\beta$ is fixed and $\rho$ varies. 
Suppose that for a given $\beta$ the equation $\zeta_c(\rho,\beta)=\rho\lambda_\beta^d$ has a unique solution for $\rho$. If $\rho\lambda_\beta^d< \zeta_c(\rho,\beta)$, all the cycles are finite and $A_{\rho,\beta}(\mu)=\rho\lambda_\beta^d$ provides $\mu=\partial{f}(\rho,\beta)/\partial\rho<{\cal E}(\rho,\beta)$. For $\rho\lambda_\beta^d>\zeta_c(\rho,\beta)$ the part $\rho-\zeta_c(\rho,\beta)/\lambda_\beta^d$ of the density is in infinite cycles, and one obtains an analytically different expression, $\partial{f}(\rho,\beta)/\partial \rho={\cal E}(\rho,\beta)$. $f(\rho,\beta)$ being a convex function of $\rho$,
$\partial f(\rho,\beta)/\partial\rho$ exists for almost all $\rho$ and is an increasing function of it. We suppose that $\partial f/\partial\rho$ is strictly increasing
and differentiable for $\rho$ close to but not equal to $\zeta_c(\rho,\beta)/\lambda_\beta^d$. Then, for $\rho< \zeta_c(\rho,\beta)/\lambda_\beta^d$
\begin{eqnarray*}
\frac{\partial^2 f(\rho,\beta)}{\partial\rho^2}
=\left(\frac{\partial\rho}{\partial\mu}\right)^{-1}_{\mu=\partial f/\partial\rho}
=\frac{\lambda_\beta^d}{\partial A_{\rho,\beta}(\mu)/\partial\mu\left|_{\mu=\partial f/\partial\rho}\right.}
=\frac{\lambda_\beta^d}{\beta}\left[\sum_{n=1}^\infty \frac{e^{n\beta[D_n(\rho,\beta)+\partial f(\rho,\beta)/\partial\rho-{\cal E}(\rho,\beta)]}}{n^{d/2-1}}\right]^{-1},
\end{eqnarray*}
and for $\rho> \zeta_c(\rho,\beta)/\lambda_\beta^d$
\[
\partial^2 f(\rho,\beta)/\partial\rho^2=\partial {\cal E}(\rho,\beta)/\partial\rho
=\partial E(\rho,\beta)/\partial\rho.
\]
The strongest singularity is obtained in three and four dimensions, if $n|D_n(\rho,\beta)|< C$ for some $C>0$. In this case $\sum_{n=1}^\infty e^{n\beta D_n(\rho,\beta)}/n^{d/2-1}=\infty$, so $\partial^2 f(\rho,\beta)/\partial\rho^2$ tends to zero when $\rho$ goes to $\zeta_c(\rho,\beta)/\lambda_\beta^d$ from below, implying the divergence of the isothermal compressibility $[\rho^2\partial^2 f(\rho,\beta)/\partial\rho^2]^{-1}$. If $\rho$ goes to $\zeta_c(\rho,\beta)/\lambda_\beta^d$ from above, $\partial^2 f(\rho,\beta)/\partial\rho^2$ attains a positive value, therefore the right limit of the compressibility is finite.
The singularity is seen also in the variation of the pressure. As a function of $\mu$ and $\beta$, the pressure is the Legendre transform of the free energy density $f(\rho,\beta)$. Substituting $\partial f(\rho,\beta)/\partial\rho$ for the chemical potential $\mu$, we obtain it as a function of $\rho$ and $\beta$,
\[
p(\rho,\beta)=\rho \frac{\partial f(\rho,\beta)}{\partial\rho} – f(\rho,\beta).
\]
The derivative of $p$ with respect to $\rho$ is therefore
\[
\frac{\partial p(\rho,\beta)}{\partial\rho} = \rho \frac{\partial^2 f(\rho,\beta)}{\partial\rho^2} \geq 0.
\]
$p(0,\beta)=0$, so $p(\rho,\beta)$ is an increasing positive function of $\rho$. From the formula for $\partial^2 f(\rho,\beta)/\partial\rho^2$ it is seen that there is a singularity at the transition point. In 3 and 4 dimensions $\partial p/\partial\rho$ is discontinuous, it decreases to zero and then jumps to a positive value when $\rho$ crosses the point where infinite cycles appear.
We expect this result for fast decaying interactions.

Interactions with a slowly decaying negative tail can give rise to the appearance of infinite cycles and an accompanying singularity in one and two dimensions. The following conjecture is based on
Eqs.~(\ref{Delta-epsilon-estimate}) and (\ref{ek+1-ek}), implying $n\delta_n(\beta)\to-\infty$, and on the assumption that the sequence $n\Delta_n(\rho,\beta)$ does not go to plus infinity at the same rate or faster.

\vspace{5pt}
\noindent
{\em Conjecture 2.} For $d=1,2$ let $u$ satisfy the bounds (\ref{upper-lower-on-u}) with $2<\eta_2\leq\eta_1<4$ and $u(x)<0$ if $|x|>R$.
Then $\zeta_c(\rho,\beta)<\infty$, for $\rho\lambda_\beta^d\leq\zeta_c(\rho,\beta)$ all the cycles are finite, and for $\rho\lambda_\beta^d>\zeta_c(\rho,\beta)$ the part $\rho-\zeta_c(\rho,\beta)/\lambda_\beta^d$ of the density is carried by infinite cycles. The analytic form of $\partial f(\rho,\beta)/\partial\rho$ changes at $\rho=\zeta_c(\rho,\beta)/\lambda_\beta^d$.

\vspace{5pt}
Assuring stability of a partly negative $u$ is nontrivial, proven only for sums of a positive-type and a nonnegative function [R].
As an example, let $v$ be a real square-integrable function, then
$
u(x)=\int v(x+y)v(y)\d y
$
is of positive type and therefore stable. Define
\begin{eqnarray*}
v(y)=
\left\{\begin{array}{ll}
v_1,& |y|<r\\
-v_2/|y|^3,&|y|>r
\end{array}
\right.
\end{eqnarray*}
with $r, v_1, v_2>0$. If $|x|>2r$, both $|y|$ and $|x+y|$ cannot be smaller than $r$; if $v_1>Cv_2/r^3$, where $C$ is large enough, then $u(x)<0$ for $|x|>2r$ and the bounds (\ref{upper-lower-on-u}) are satisfied with $\eta_1=\eta_2=3$. ($C=11$ works for $d=1$.)

\section{Concluding remarks}

The transition from finite to infinite cycles is more fundamental than BEC and should be considered as a phase transition in its own right. It is reminiscent of percolation transition,
except for the fact that on lattices the infinite cluster is unique while in the Bose problem we expect infinitely many infinite cycles [S2]. From this point of view the analogy with percolation on the infinite homogeneous Caley tree is closer, on the tree the number of infinite clusters is infinite [FE].
The percolation transition is known to give rise to a weak singularity in the analogue of the free energy at the percolation threshold [KS]. The singularity we found is much stronger; notably, in three and four dimensions the isothermal compressibility presents a one-sided divergence or, equivalently, the $\rho$-derivative of the pressure is discontinuous at the transition point. We recall that in the ideal gas $\partial f(\rho,\beta)/\partial\rho\equiv 0$ for $\rho>\zeta(d/2)/\lambda_\beta^d$, and $\partial p(\rho,\beta)/\partial\rho=\rho\partial^2 f(\rho,\beta)/\partial\rho^2$ is continuous with zero value at $\rho=\zeta(d/2)/\lambda_\beta^d$.

It has long been known that interactions weaken BEC by scattering particles out of the zero-momentum state . It is somewhat surprising, however, that long-range repulsion can cause the complete disappearance of infinite cycles, and hence of BEC, in three dimensions. Our argument predicts this to happen only for $d=3$, and the survival of infinite cycles in $d\geq 4$ dimensions is probably not an artefact of our approach.
Also surprising, long-range attraction can lead to the appearance of infinite cycles in one and two dimensions, certainly without BEC [H]. However, it was not entirely unexpected that some transition would take place at high densities or low temperatures, knowing that ordinary condensation is possible even in one dimension for classical particles interacting via Lennard-Jones type pair potentials [J].

If the critical value $\zeta_c(\rho,\beta)$ of $\rho\lambda_\beta^d$ remains finite in the limit $\beta\to\infty$, at zero temperature all the particles are in infinite cycles. Because $\beta\delta_n(\beta)\to 0$ in this limit (at least for $u\geq 0$), $\zeta_c(\rho,\beta)\to\infty$ could come only from $\beta\Delta_n(\rho,\beta)\to \infty$ for some values of $n$. At a given $\rho$, finite and infinite cycles can coexist in the ground state only if
\[
0<\lim_{\beta\to\infty}\zeta_c(\rho,\beta)/\rho\lambda_\beta^d<1,
\]
which supposes $\Delta_n(\rho,\beta)\sim\ln\rho\lambda_\beta^d/n\beta$ when $\beta$ goes to infinity. However, we consider $\Delta_n(\rho,\beta)\equiv 0$ more probable, with the consequence that there are only infinite cycles at zero temperature. If that is true, combining with mathematically rigorous [LY], numerical [C] and experimental [SS] results, all showing that the ground state of the interacting gas is not fully Bose-condensed, one concludes that infinite cycles can carry other than the condensate. This is particularly interesting in view of the fact that superfluidity is 100$\%$ but BEC is less than 10$\%$ in the ground state of the helium liquid [PO], [C], [SS]. It suggests that superfluidity can be related to infinite cycles rather than BEC. 2D superfluidity without BEC points in the same direction.

We do not have a firm result about the sign of the shift of the critical temperature, relative to the ideal gas. According to the consensus, in the case of finite-range repulsive interactions at low temperatures the transition temperature should be shifted upward [BBHLV], [B], [SU].  At least, the Lemma does not contradict this by allowing ${\cal E}_{n+1}- {\cal E}_n< 0$ for small $n$ and $\zeta_c(\rho,\beta)<\zeta(d/2)$ as a consequence.

An intuitive understanding of the transition from finite to infinite cycles is possible through an energy-entropy argument.
Transferring density into long cycles decreases the entropy $-\sum_{n=1}^N (\rho^{N,L}_n/\rho)\ln(\rho^{N,L}_n/\rho)$, but the decrease of energy can be more important. A cycle of length $n$ is an effective one-particle trajectory with $t$ running from 0 to $n\beta$. According to a natural interpretation [S3], the $n$ particles composing it are in the same one-particle state that spreads over a sphere of radius $\sim\sqrt{n}\lambda_{\beta}$. Thus, the particles in long cycles are in extended states, the overlap of their wave function with those of other particles is small everywhere, decreasing thereby the energy per particle and the free energy at large densities. In a probability versus density plot, for $\rho$ small $\rho^{N,L}_n/\rho$ has a single maximum at $n$ of order~1, dividing into two and producing a second maximum at $n\gg 1$, when $\rho$ is large.
In three dimensions the long-range repulsion can exceed a critical strength that causes the advantage of forming long cycles disappear. Long-range attraction can have the opposite effect, giving rise to infinite cycles also in one and two dimensions.

\vspace{10pt}\noindent
{\bf Acknowledgement.} This work was supported by the Hungarian Scientific Research Fund (OTKA) through Grant No. K146736.



\vspace{10pt}
\noindent
{\Large\bf References}

\begin{enumerate}
\item[{[AD1]}] S. Adams and M. Dickson, {\em An explicit large deviations analysis of the spatial cycle Huang-Yang-Luttinger model.} Ann. Henri Poincar\'e {\bf 22}, 1535-1560 (2021).
\item[{[AD2]}] S. Adams and M. Dickson, {\em Large deviations analysis for random combinatorial partitions with counter terms.} J. Phys. A: Math. Theor. {\bf 55}, 255001 (2022).
\item[{[BBHLV]}] G. Baym, J.-P. Blaizot, M. Holzmann, F. Laloë, and D. Vautherin, {\em Bose-Einstein transition in a dilute interacting gas.} Eur. Phys. J. B {\bf 24}, 107-124 (2001).
\item[{[BCMP]}] Benfatto G., Cassandro M., Merola I. and Presutti E.: {\em Limit theorems for statistics of combinatorial partitions with applications to mean field Bose gas.} J. Math. Phys. {\bf 46}, 033303 (2005).
\item[{[BU1]}] V. Betz and D. Ueltschi, {\em Spatial random permutations and infinite cycles.} Commun. Math. Phys. {\bf 285}, 469-501 (2009).
\item[{[BU2]}] V. Betz and D. Ueltschi, {\em Spatial random permutations and Poisson-Dirichlet law of cycle lengths.} Electr. J. Probab. {\bf 16}, 1173-1192 (2011).
\item[{[BUV]}] V. Betz, D. Ueltschi and Y. Velenik, {\em Random permutations with cycle weights.} Ann. Appl. Probab. {\bf 21}, 312-331 (2011).
\item[{[Bl]}] J. Blaizot, {\em Non Perturbative Renormalization Group and Bose-Einstein Condensation.} arXiv:0801.0009
\item[{[Bo]}] G. Boland, {\em Long cycles in the infinite-range-hopping Bose–Hubbard model.} J. Math. Phys. {\bf 50}, 073301 (2009).
\item[{[BP]}] G. Boland and J.V. Pulé, {\em Long Cycles in the Infinite-Range-Hopping Bose-Hubbard Model with Hard Cores.} J. Stat. Phys. {\bf 132}, 881–905  (2008).
\item[{[C]}] D. M. Ceperley, {\em Path integrals in the theory of condensed helium.} Rev. Mod. Phys. {\bf 67}, 279-355 (1995).
\item[{[DMP]}]  T. C. Dorlas, Ph. A. Martin and J. V. Pul\'e, {\em Long cycles in a perturbed mean field model of a boson gas.} J. Stat. Phys. {\bf 121}, 433-461 (2005).
\item[{[DV]}] M. Dickson and Q. Vogel, {\em Formation of infinite loops for an interacting bosonic loop soup.} Electron. J. Probab. {\bf 29}, 1-39 (2024).
\item[{[D]}] F. J. Dyson, {\em Existence of a Phase-Transition in a One-Dimensional Ising Ferromagnet.} Commun. Math. Phys. {\bf 12}, 91-107 (1969).
\item[{[EP]}] Elboim D. and Peled R.: {\em Limit distributions for Euclidean random permutations.} Commun. Math. Phys. {\bf 369}, 457-522 (2019).
\item[{[F]}] R. P. Feynman, {\em Atomic theory of the $\lambda$ transition in helium.} Phys. Rev. {\bf 91}, 1291-1301 (1953).
\item[{[FE]}]  M. E. Fisher and J. W. Essam, {\em Some cluster size and percolation problems.} J. Math. Phys. {\bf 2}, 609-619 (1961).
\item[{[G]}] J. Ginibre, {\em Some applications of functional integration in Statistical Mechanics.} In: {\em Statistical Mechanics and Quantum Field Theory}, eds. C. De Witt and R. Stora, Gordon and Breach (New York 1971).
\item[{[H]}] P. C. Hohenberg, {\em Existence of long-range order in one and two dimensions.} Phys. Rev. {\bf 158}, 383-386 (1967).
\item[{[J]}] K. Johansson, {\em On Separation of Phases in One-Dimensional Gases.} Commun. Math. Phys. {\bf 169}, 521-561 (1995).
\item[{[KU]}] B. Kahn and G. E. Uhlenbeck, {\em On the theory of condensation.} Physica {\bf 5}, 399-416 (1938).
\item[{[KR]}] M. G. Krein and M. A. Rutman, Am. Math. Soc. Transl. Series 1 {\bf 10}, 199 (1962) [Usp. Mat. Nauk. {\bf 3}, 3 (1948).
\item[{[KS]}] H. Kunz and B. Souillard, {\em Essential Singularity in Percolation Problems and
Asymptotic Behavior of Cluster Size Distribution.} J. Stat. Phys, {\bf 19}, 77-106 (1978).
\item[{[LY]}] E. H. Lieb and J. Yngvason, {\em Ground state energy of the low density Bose gas.} Phys. Rev. Lett. {\bf 80}, 2504-2507 (1998).
\item[{[L]}] F. London, {\em On the Bose-Einstein condensation.} Phys. Rev. {\bf 54}, 947-954 (1938).
\item[{[LHB]}] J. L\H orinczi, F. Hiroshima and V. Betz, {\em Feynman-Kac-Type Theorems and Gibbs Measures on Path Space.} De Gruyter, Berlin/Boston (2011).
\item[{[PO]}] O. Penrose and L. Onsager, {\em Bose-Einstein condensation and liquid He.} Phys. Rev. {\bf 104}, 576-584 (1956).
\item[{[Ro]}] G. Roepstorff, {\em Path Integral Approach to Quantum Physics.} Springer-Verlag 1994.
\item[{[R]}] D. Ruelle,. \emph{Statistical Mechanics.} W. A. Benjamin (New York-Amsterdam 1969).
\item[{[SU]}] R. Seiringer and D. Ueltschi, {\em Rigorous upper bound on the critical temperature of dilute Bose gases.} Phys. Rev. B {\bf 80}, 014502 (2009).
\item[{[Si]}] B. Simon, {\em Functional Integration and Quantum Physics.} AMS Chelsea Publ. 2005.
\item[{[SS]}] W. M. Snow and P. E. Sokol, {\em Density and temperature dependence of the momentum distribution in liquid Helium 4.} J. Low. Temp. Phys. {\bf 101}, 881-928 (1995).
\item[{[S1]}]  A. S\"ut\H o, {\em Percolation transition in the Bose gas.} J. Phys. A: Math. Gen. {\bf 26}, 4689-4710 (1993).
\item[{[S2]}] A. S\"ut\H o, {\em Percolation transition in the Bose gas: II.} J. Phys. A: Math. Gen. {\bf 35}, 6995-7002 (2002). See also arXiv:cond-mat/0204430v4 with addenda after Eqs.~(34) and (44).
\item[{[S3]}] A. S\"ut\H o, arXiv:2305.18959 [math-ph].
\item[{[Tth]}] T\'oth B.: {\em Phase transition in an interacting Bose system. An application of the theory of Ventsel’ and Freidlin.} J. Stat. Phys {\bf 61}, 749–764 (1990).
\item[{[U1]}] D. Ueltschi, {\em Relation between Feynman cycles and off-diagonal long-range order.} Phys. Rev. Lett. {\bf 97}, 170601 (2006).
\item[{[U2]}] D. Ueltschi, {\em Feynman cycles in the Bose gas.} J. Math. Phys. {\bf 47}, 123303 (2006).
\item[{[Uh]}] G. E. Uhlenbeck, Dissertation Leiden, 1927, p. 69.

\end{enumerate}

\end{document}